\documentstyle[12pt]{article}
\begin{document}

\input mssymb.tex
\def\ov{\overline}
\def\var{\varepsilon}
\def\om{\omega}
\def\pa{\partial}
\def\la{\langle}
\def\ra{\rangle}
\def\om{\omega}

\title{The stochastic limit \\ of quantum spin systems}
\author{L. Accardi, S.V. Kozyrev}
\maketitle
\centerline{\it Centro Vito Volterra Universita di Roma Tor Vergata}

\begin{abstract}
The stochastic limit for the system of spins
interacting with a boson field is investigated.
In the finite volume an application of the stochastic golden rule
shows that in the limit the dynamics of a quantum system
is described by a quantum white noise equation that after taking
of normal order is equivalent to quantum stochastic differential
equation (QSDE).
For the quantum Langevin equation the dynamics is well defined
and is a quantum flow on the infinite lattice system.
\end{abstract}

\section{Introduction}

Starting with the work of Glauber \cite{Glauber}
the dynamics of infinite classical lattice systems has been considered
by many authors and has led to study the
ergodic and equlibrium properties of a new class of
classical Markov semigroups (cf. \cite{Liggett} for a general survey and for
further references). Quantum analogues of these semigroups have also
been considered by several authors (e.g. \cite{Martin}, \cite{Matsui},
\cite{MajZeg96a}, \cite{Spitzer}, \cite{Sullivan}, \dots).
However the problem of deriving these Markovian
semigroups, and more generally the stochastic flows, as limits of
Hamiltonian systems, was open both in the classical and in the quantum case.

On the other hand the stochastic golden rules, which arise in the
stochastic limit of quantum theory as natural generalizations of Fermi
golden rule \cite{AcLuVo}, provide a natural tool to associate a
stochastic flow, driven by a white noise equation, to any discrete
system interacting with a quantum field. Moreover another quite general
result of the stochastic limit is that the Markov semigroup,
canonically associated to the flow,
leaves invariant the abelian subalgebra generated by the spectral
projections of the Hamiltonian of the discrete system so that the associated
Markov process describes the jumps among these energy levels.

Since a quantum spin system in finite volume is obviously a discrete
system, the above results naturally suggest the conjecture that,
by coupling such a system to a quantum field via a suitable interaction,
applying the stochastic golden rule and taking the thermodynamic limit,
one might obtain a class of quantum flows which, when restricted to an
appropriate abelian subalgebra, gives rise to the classical interacting
particle systems studied in classical statistical mechanics.

In the present paper we will prove this conjecture for general
finite range interactions thus extending a previous result obtained
by the authors in the case of the Ising model \cite{AcKo99a}.
The model we consider is essentially the
same as the one considered by Martin and Buffet \cite{Martin} apart from
the minor difference that these authors consider fermion reservoir and
we a boson one. Since the stochastic golden rule holds also in the
fermion case \cite{AcFriLu}, \cite{AcLuVo} there is no difficulty in
extending the present results to the fermion case.
We investigate a quantum system with a Hamiltonian of the form
$$
H=H_0 +\lambda H_I
$$
where $H_0$ is called the free Hamiltonian and $H_I$ the
interaction Hamiltonian.
The general idea of the stochastic limit
(see \cite{AcLuVo99}) is to make the time rescaling
$t\to t/\lambda^2$ in the solution of the Schr\"o\-din\-ger
(or Heisenberg) equation in
interaction picture $U^{(\lambda)}_t=e^{itH_0} e^{-itH}$,
associated to the Hamiltonian $H$, i.e.
$$
{ \frac{\partial}{\partial t} }  U^{( \lambda)}_t
=-i { \lambda  } H_I (t) \ U_t^{(\lambda )}
$$
with $H_I(t)=e^{it H_0}H_Ie^{-itH_0}$.
This gives the rescaled equation
\begin{equation}\label{AK_resc_evol}
{ \frac{\partial}{  \partial t} }  U^{( \lambda )}_{t/\lambda^2}
=- {\frac{i}{\lambda}} H_I (t/\lambda^2) \
U_{t/\lambda^2}^{(\lambda)}
\end{equation}
and one wants to study the limits, in a topology to be specified,
\begin{equation}\label{AK_1.6}
\lim_{\lambda\to0}U^{(\lambda)}_{t/\lambda^2}= U_t;\qquad
\lim_{\lambda\to0}{\frac{1}{\lambda}}\,H_I\left({\frac{t}{\lambda^2}}\right)
=H_t
\end{equation}
After the rescaling $t\to t/\lambda^2$ we consider the
simultaneous limit $\lambda\to 0$, $t\to\infty$ under the condition
that $\lambda^2 t$ tends to a constant (interpreted as a new slow time
scale). This limit captures the dominating
contributions to the dynamics, in a regime of long times and small
coupling, arising  from the cumulative effects, on a large time scale,
of small interactions ($\lambda\to 0$). The physical idea is that,
looked from the slow time scale of the atom, the field looks like a very
chaotic object: a {\it quantum white noise}, i.e. a $\delta$-correlated
(in time) quantum field $b^{*}(t,k), b(t,k)$ also called a {\it master
field}.

\section{Physical model}

In the present paper we consider the open Ising model, describing the
interaction of a system $S$ of spins (or, more generally, two-level systems)
with a reservoir, represented by a bosonic quantum field.
The total Hamiltonian is
$$
H=H_0+\lambda H_I=H_S+H_R+\lambda H_I
$$
where $H_R$ is the free Hamiltonian of a bosonic reservoir $R$:
$$
H_R=\int\omega(k)a^{*}(k) a(k)\,dk
$$
acting in the representation space ${\cal F}$ corresponding to
the bosonic equilibrium state at temperature $\beta^{-1}$ and chemical
potential $\mu$ and the assumptions we need on $\omega(k)$ are given at
the beginning of Section III below. Thus the reservoir state is
Gaussian with mean zero and correlations given by
$$
\langle a^{*}(k)a(k')\rangle=
{1\over{e^{\beta(\omega(k)-\mu)}-1}}\delta(k-k')
$$
The spin variables are labeled by the lattice $Z^d$, and,
for each finite subset $\Lambda\subseteq Z^d$,
the system Hilbert space is
$$
{\cal H}_S={\cal H}_{\Lambda}=\otimes_{r\in\Lambda} C^2
$$
and the system Hamiltonian has the form
$$
H_S=H_{\Lambda}=-{1\over2}\,\sum_{r,s\in\Lambda}J_{rs}\sigma^z_r\sigma^z_s
$$
where $\sigma^x_r$, $\sigma^y_r$, $\sigma^z_r$ are Pauli matrices
$(r\in\Lambda)$ at the $r$-th site in the tensor product.
For any $r$, $s\in\Lambda$
$$
J_{rs}=J_{sr}\in R, \qquad J_{rr}=0
$$
In the present paper we consider a system Hamiltonian that
describes the interaction of spin with a finite number of other spins
(finite range potential).
The simplest example is the nearest neighbor interaction
(Ising model), discussed in Section V below.
The interaction  Hamiltonian $H_I$
(acting in ${\cal H}_S\otimes {\cal F}$) has the form
$$
H_I=\sum_{r\in \Lambda}\sigma^x_r\otimes\psi(g_r), \quad
\psi(g)=A(g)+A^{*}(g),\quad A(g)=\int dk \, \overline{g}(k)a(k),
$$
where $\psi$ is a field operator, $A(g)$ is a smeared quantum field with
cutoff function (form factor) $g(k)$.
To perform the construction of the stochastic limit one needs to calculate
the free evolution of the interaction Hamiltonian:
$H_I(t)=e^{itH_0}H_I  e^{-itH_0}$.
We use the formula
$$
\sigma^z_r=
\sum_{\varepsilon_r=\pm1}\varepsilon_r|\varepsilon_r\rangle
\langle \varepsilon_r|
$$
where $|\varepsilon_r\rangle$ is the eigenvector of $\sigma^z_r$
corresponding to the egenvalue $\varepsilon_r$
to rewrite the interaction Hamiltonian in the form
$$
H_I=\sum_{r\in\Lambda}\sum_{\var_r,\mu_r=\pm1}
\langle\var_r,\sigma^x_r\mu_r\rangle
|\var_r\rangle\langle\mu_r|\otimes(A(g)+A^{*}(g))=
\left(D_\Lambda+D^{*}_\Lambda\right)\left(A(g)+A^{*}(g)\right)
$$
where
$$
D_r=|1_r\rangle\langle -1_r|\quad ,\quad D_r^*=|-1_r\rangle\langle 1_r|
\quad ,\quad D_\Lambda=\sum_{r\in\Lambda}D_r
$$
Since
$$
[H_S,|1_r\rangle\langle {-1}_r|]=
\Delta(r)|1_r\rangle\langle {-1}_r|\qquad, \qquad
\Delta(r)=-\sum_{s}J_{rs}\sigma^z_{s}
$$
the free evolution for $D_r$ has the form
$$
e^{itH_S}D_r e^{-itH_S}=
e^{it\Delta(r)}D_r
$$
The sum over $s$ here is finite because the spin at $r$ interacts
only with a finite number of spins (only a finite number of $J_{rs}$
for fixed $r$ is nonzero).

Using the formula for the free evolution of bosonic fields
$$
e^{itH_R}a(k)e^{-itH_R}=e^{-it\omega(k)}a(k)
$$
we get for the free evolution of the interaction Hamiltonian:
\begin{equation}\label{AK_evolvedih}
H_I(t)=
\sum_{r\in\Lambda}\int dk \,\overline{g}(k)\left(
D_r e^{-{it}(\omega(k)-\Delta(r))}a(k)+
D_r^{*} e^{-{it}(\omega(k)+\Delta(r))}a(k)
\right)+h.c.
\end{equation}
and, for the free evolution of $D_r$:
$$
e^{it\Delta(r)}D_r=e^{-it\sum_{s\in\Sigma(r)}J_{rs}\sigma^z_{s}}D_r=
D_r\prod_{s\in\Sigma(r)}e^{-itJ_{rs}\sigma^z_{s}}=
$$
\begin{equation}\label{AK_fordr}
=D_r\prod_{s\in\Sigma(r)}\sum_{\varepsilon_s=\pm1}
e^{-itJ_{rs}\varepsilon_s}
|\varepsilon_s\rangle \langle \varepsilon_s|
\end{equation}
where $\Sigma(r)$ denotes the (finite) set of spins interacting with
the spin $r$ (so $r\ne\Sigma(r)$).
When no confusion is possible we use the same symbol $\Sigma(r)$
to denote also the set of configurations of correponding spins:
$\sigma(r)=\{\varepsilon_s \} (s\in\Sigma(r))$.

We denote
\begin{equation}\label{AK_energy}
E(r,\sigma(r))=\sum_{s\in\Sigma(r)}J_{rs}\varepsilon_s=
H_S(\sigma_r^z=-1,\sigma(r))-H_S(\sigma_r^z=-1,\sigma(r))
\end{equation}
the energy difference between two configurations
with all spins fixed, and given by $\sigma(r)$, with the
exception of $\varepsilon_r$ that changes value from $-1$ to $1$.

With these notations formula (\ref{AK_evolvedih}) takes the form
$$
H_I(t)=\sum_{r\in\Lambda}\int dk \,\overline{g}(k)\biggl(
\sum_{\sigma\in\Sigma(r)}
D_r\prod_{s\in\Sigma(r)}
|\varepsilon_s\rangle \langle \varepsilon_s|
e^{-it\left(\omega(k)+E(r,\sigma(r))\right)}a(k)+
$$
$$
+\sum_{\sigma\in\Sigma(r)}D_r^{*}\prod_{s\in\Sigma(r)}
|\varepsilon_s\rangle \langle \varepsilon_s|
e^{-it\left(\omega(k)-E(r,\sigma(r))\right)}a(k)
\biggr)+h.c.=
$$
$$
=\sum_{r\in\Lambda}\sum_{\sigma(r)\in\Sigma(r)}
\int dk \,\overline{g}(k)\biggl(
G_{r,\sigma(r)}
e^{-it\left(\omega(k)+E(r,\sigma(r))\right)}a(k)+
$$
$$
+G_{r,\sigma(r)}^{*}
e^{-it\left(\omega(k)-E(r,\sigma(r))\right)}a(k)
\biggr)+h.c. =
$$
\begin{equation}\label{AK_fullint}
=\sum_{r\in\Lambda}\sum_{\sigma(r)\in\Sigma(r)}
\int dk \,\overline{g}(k)F_{r,\sigma(r)}^{*}
e^{-it\left(\omega(k)-E(r,\sigma(r))\right)}a(k) + h.c.
\end{equation}
where we denote
\begin{equation}\label{AK_F}
G_{r,\sigma(r)}=
D_r\prod_{s\in\Sigma(r)}
|\varepsilon_s\rangle \langle \varepsilon_s|,\qquad
F_{r,\sigma(r)}=G_{r,-\sigma(r)}^{*}+G_{r,\sigma(r)}
\end{equation}
To prove (\ref{AK_fullint}) we used a suitable rearranging of the indices
$\sigma(r)$ and the property
$$
E(r,-\sigma(r))=-E(r,\sigma(r))
$$
where $-\sigma(r)$ is the configuration of spins with all spins
opposite to the spins in $\sigma(r)$, i.e.
$\varepsilon_s(-\sigma(r))=-\varepsilon_s(\sigma(r))$.

The operator $F_{r,\sigma(r)}$
flips the spin at site $r$ (or kills a spin configuration).


\section{The stochastic limit of the model}

In this section we will denote $R\in {\cal S}$ the pair of indices
$(r,\sigma(r))$ where $r$ takes values in $\Lambda$. In these notations
the free evolution of the interaction Hamiltonian (\ref{AK_fullint}) takes
the form
\begin{equation}\label{AK_tiih}
H_I(t)=\sum_{R\in {\cal S}}\int dk \,\overline{g}(k)F_{R}^{*}
e^{-it\left(\omega(k)-E(R)\right)}a(k) + h.c.
\end{equation}

In the stochastic limit the field $H_I(t)$ gives rise to
a family of quantum white noises, or master fields.
To investigate these noises, let us suppose the following:

1) $\omega(k)\ge 0,\qquad \forall k$;

2) The $d-1$--dimensional Lebesgue measure of the surface
$\{k:\omega(k)=0\}$ is equal to zero (so that $\delta(\omega(k))=0$)
(for example $\omega(k) = k^2 + m$ with $m\geq 0$).

Now let us investigate the limit of $H_I(t/\lambda^2)$ using one of the the
basic formulae of the stochastic limit:
\begin{equation}\label{AK_imformula}
\lim_{\lambda\to 0} {1\over\lambda^2} \exp\left({it\over\lambda^2}f(k)\right)=
2\pi \delta(t)\delta(f(k))
\end{equation}
Since the term $\delta(f(k))$ in (\ref{AK_imformula}) is not identically equal
to zero only if $f(k)=0$ for some $k$ in a set of nonzero $d-1$--dimensional
Lebesgue measure (in our case $f(k)=\omega(k)-E(R)$),
in the stochastic limit of the rescaled interaction
Hamiltonian (\ref{AK_tiih}) only the terms with $R\in {\cal S_+}$ will
survive. Here ${\cal S_+}$ (respectively ${\cal S_-}$)
denotes the spin configurations $(\varepsilon_s)$
whose mean energy difference $\sum_{s\in\Sigma(r)}J_{rs}\varepsilon_s$
is strictly positive (respectively negative).
We will call such spin configurations positive (respectively negative).

The rescaled interaction in (\ref{AK_tiih}) is expressed in terms of the
rescaled creation and annihilation operators
$a_{\lambda,R}(t,k)={1\over\lambda}\,
e^{-i{t\over\lambda^2}(\omega(k) - E(R))}a(k)$.
After the stochastic limit every rescaled annihilation operator
correspondent to positive spin configuration
generates one non-trivial quantum white noise
$$
b_{R}(t,k)=\lim_{\lambda\to 0}a_{\lambda,R}(t,k)=
\lim_{\lambda\to 0}{1\over\lambda}\,
e^{-i{t\over\lambda^2}(\omega(k)-E(R))} a(k)
$$
with the relations
$$
[b_{R}(t,k), b_{R}^{*}(t',k')]=
\lim_{\lambda\to 0}[a_{\lambda,R}(t,k),
a^{*}_{\lambda,R}(t',k')]=
$$
\begin{equation}\label{AK_noiserelat}
=
\lim_{\lambda\to 0}{1\over\lambda^2}\,
e^{-i{t-t'\over\lambda^2}(\omega(k)-E(R))}\delta(k-k')=
2\pi\delta(t-t')\delta(\omega(k)-E(R)) \delta(k-k')
\end{equation}
$$
[b_{R}(t,k), b_{R'}^{*}(t',k')]=0
$$
by (\ref{AK_imformula}). Moreover, configurations, corresponding
to different values of the energy difference $E(R)$, are independent.
For generic interactions if $R\ne R'$ then corresponding energy
differences $E(R)\ne E(R')$ and correponding white noises are independent.

The stochastic limit of the interaction Hamiltonian
is therefore equal to
\begin{equation}\label{AK_h}
h(t)=\lim_{\lambda\to 0}
{1\over\lambda}\,H_I\left({t\over\lambda^2}\right)=
\int dk \,\overline{g}(k) \sum_{R\in {\cal S_+}}
F^{*}_{R} b_{R}(t,k)+h.c.
\end{equation}
Physically such a form of the evolved interaction Hamiltonian
can be explained as follows: the operator
$F_{R}$, where $R\in {\cal S_+}$,
decreases the energy of the spin configuration.
Therefore the vertex $F^{*}_{R} a(k)$ describes the absorption, by the spin
configuration $R$, of an energy quantum of the field of momentum $k$.

The state on the master field (white noise) $b_{R}(t,k)$, corresponding
to the equilibrium state of the field, is the mean zero gaussian state with
correlations:
$$
\langle  b^{*}_{R}(t,k) b_{R}(t',k')\rangle=
2\pi\delta(t-t')\delta(\omega(k)-E(R))\delta(k-k')
{1\over{e^{\beta(\omega(k)-\mu)}-1}}
$$
$$
\langle  b_{R}(t,k) b^{*}_{R}(t',k')\rangle=
2\pi\delta(t-t')\delta(\omega(k)-E(R))\delta(k-k')
{1\over{1-e^{-\beta(\omega(k)-\mu)}}}
$$
and vanishes for noises corresponding to different positive spin
configurations.

Now let us investigate the evolution equation in interaction picture for
our model. According to the general scheme of the stochastic limit,
up to possible divergences (due to the thermodynamic limit) that we will
discuss later, we get the (singular) white noise equation
\begin{equation}\label{AK_noiseequ}
{d\over dt}U_{t}= -i h(t) U_t
\end{equation}
whose normally ordered form is the quantum stochastic differential equation
\cite{AcLuVo99b}
\begin{equation}\label{AK_QSDE}
dU_{t}=\left(-i dH(t)-  Gdt\right)U_t
\end{equation}
where $h(t)$ is the white noise (\ref{AK_h}) given by the stochastic limit
of the interaction Hamiltonian and
\begin{equation}\label{AK_dH}
dH(t)=\sum_{R\in {\cal S_+}}
\left(F^{*}_{R} dB_{R}(t)+F_{R}dB^{*}_{r,\sigma(r)}(t)\right)
\end{equation}
\begin{equation}\label{AK_dB}
dB_{R}(t)=\int dk \,\overline{g}(k)
\int_t^{t+dt}b_{R}(\tau,k) d\tau
\end{equation}

According to the stochastic golden rule, (\ref{AK_QSDE}) is obtained as
follows: the first term in (\ref{AK_QSDE}) is just the limit of the
iterated series solution for (\ref{AK_resc_evol})
$$
\lim_{\lambda\to 0} {1\over \lambda}\,
\int_{t}^{t+dt}H_I\left({\tau\over \lambda^2}\right)d\tau
$$
The second term $Gdt$, called the drift, is equal to the limit
of the expectation value in the reservoir state of the second term
in the iterated series solution for (\ref{AK_resc_evol})
$$
\lim_{\lambda\to 0} {1\over \lambda^2}\,
\int_{t}^{t+dt}dt_1\, \int_{t}^{t_1}dt_2\, \langle
H_I\left({t_1\over \lambda^2}\right) H_I\left({t_2\over \lambda^2}\right)
\rangle$$
Making in this formula the change of variables $\tau=t_2-t_1$ we get
\begin{equation}\label{AK_term2}
\lim_{\lambda\to 0} {1\over \lambda^2}\,
\int_{t}^{t+dt}dt_1\, \int_{t-t_1}^{0}d\tau\,
\langle H_I\left({t_1\over \lambda^2}\right)
H_I\left({t_1\over \lambda^2}+{\tau\over \lambda^2}\right)\rangle
\end{equation}
Computing the (Gaussian) expectation value and
using formula (\ref{AK_tiih}) and the fact that the limits of oscillating
factors of the form  $\lim_{\lambda\to 0}e^{{ict_1\over \lambda^2}}$
vanish unless the constant $c$ is equal to zero, we see that we
can have non--zero limit only when all oscillating factors
of a kind $e^{{ict_1\over \lambda^2}}$ (with $t_1$)
in (\ref{AK_term2}) cancel. In conclusion we get
$$
G = \sum_{R\in {\cal S}}\int dk\, |{g}(k)|^2
\int^{0}_{-\infty}d\tau
\biggl(
F_{R}^{*}F_{R} e^{i{\tau}(\omega(k)-E(R))}
{1\over{1-e^{-\beta(\omega(k)-\mu)}}}+
$$
$$
+F_{R}F_{R}^{*} e^{-i{\tau}(\omega(k)-E(R))}
{ {1}\over{e^{\beta(\omega(k)-\mu)}-1} }
\biggr)
$$
and therefore, from the formula
\begin{equation}\label{AK_imformula2}
\int_{-\infty}^{0} e^{it\omega}={-i\over{\omega-i0}}=
\pi\delta(\omega)-iP.P.{1\over\omega}
\end{equation}
we get the following expression for the drift $G$:
$$
\sum_{R\in {\cal S}}\int dk\, |{g}(k)|^2
\biggl(
{-i F_{R}^* F_{R} \over{\omega(k)-E(R)-i0}}
{1\over{1-e^{-\beta(\omega(k)-\mu)}}}+
$$
$$
+{i F_{R} F_{R}^{*} \over{\omega(k)-E(R)+i0}}
{1\over{e^{\beta(\omega(k)-\mu)}-1}}
\biggr)
=$$
\begin{equation}\label{AK_drift}
=\sum_{R\in {\cal S}}\left(
F_{R}^{*} F_{R}(g|g)^-_{R}+
F_{R} F_{R}^{*}\overline{(g|g)}^+_{R}\right)
\end{equation}
We see from formula (\ref{AK_drift}) that our model also exhibits the
{\it Cheshire cat effect} discovered in \cite{AcKoVol} in a simpler
model, that is: the part of the terms in the drift (\ref{AK_drift})
with $R\in {\cal S_+}$ comes from quantum white noises (or, correspondingly,
from stochastic differentials) and describes the self--interaction of such
noises, but products of pairs of operators $F_{R}^{\varepsilon}$
with the index $R$ corresponding to negative spin configuration
describes self--interaction of virtual noises (corresponding to vertices
without conservation of energy).

The drift term contains sums over $R$ which are divergent
for large $\Lambda$. Therefore in the Schrodinger picture we will get a
divergence in the thermodynamic limit. In the following section we will
consider the evolution in the Heisenberg picture (the Langevin equation)
and we will show that, in this context, no divergence arises.

\section{The Langevin equation}

Now we will find the Langevin equation, which is the limit of
the Heisenberg evolution, in interaction representation,  of a system
observable. Let $X$ be a local operator acting on the spin degrees of
freedom, i.e. one acting only on a finite number of spins.
The Langevin equation is the equation satisfied by the stochastic flow
$j_t$, defined by:
$$
j_t\left(X\right)=U^{*}_t X U_t
$$
where $U_t$ satisfies equation (\ref{AK_QSDE}) in the previous section,
i.e.
\begin{equation}\label{AK_dU}
dU_t=\left(-idH(t)-Gdt\right)U_t
\end{equation}
\begin{equation}\label{AK_G}
G=
\sum_{R\in {\cal S}}
\left(F_{R}^{*} F_{R}(g|g)^-_{R}+
F_{R} F_{R}^{*}\overline{(g|g)}^+_{R}\right)
\end{equation}
To derive the Langevin equation we consider
\begin{equation}\label{AK_deriveL}
dj_t\left(X\right)=
j_{t+dt}\left(X\right)-j_{t}\left(X\right)=
dU^{*}_t X U_t + U^{*}_t X dU_t + dU^{*}_t X dU_t
\end{equation}
The only nonvanishing products for quantum stochastic differentials are
\begin{equation}\label{AK_products}
dB_{R}(t)dB^{*}_{R}(t)= 2\hbox{Re}\,(g|g)^{-}_{R} dt,\qquad
dB^{*}_{R}(t)dB_{R}(t)= 2\hbox{Re}\,(g|g)^{+}_{R} dt.
\end{equation}
Combining terms in (\ref{AK_deriveL}) and using (\ref{AK_dU}), (\ref{AK_dH}),
(\ref{AK_G}) and (\ref{AK_products}) we get
the Langevin equation
\begin{equation}\label{AK_Langevin} dj_t(X)
=\sum_{\alpha}
j_t\circ \theta_{\alpha}(X) dM^{\alpha}(t)
=\sum_{n=-1,1;R\in {\cal S_+}}
j_t\circ \theta_{nR}(X) dM^{nR}(t)+
j_t\circ \theta_0(X)dt
\end{equation}
where
\begin{equation}\label{AK_minus1}
dM^{-1,R}(t)=dB_{R}(t),\quad \theta_{-1,R}(X)=-i[X,F_{R}^{*}],
\quad R\in {\cal S_+}
\end{equation}
\begin{equation}\label{AK_plus1}
dM^{1,R}(t)=dB^{*}_{R}(t),\quad \theta_{1,R}(X) =-i [X,F_{R}],
\quad R\in {\cal S_+}
\end{equation}
and
\begin{equation}\label{AK_zero}
\theta_0(X)=
\sum_{R\in {\cal S}}
\left(
\theta_{0,R}^{(0,-1)}+
\theta_{0,R}^{(0,1)}+
\theta_{0,R}^{(-1)}+
\theta_{0,R}^{(1)}
\right)(X)=
\end{equation}
$$
=\sum_{R\in {\cal S}}\biggl(
-i \hbox{Im} \,(g|g)^-_R [X, F_{R}^* F_{R} ]
+i \hbox{Im} \,(g|g)^+_R [X, F_{R} F_{R}^{*}]
+
$$
$$
+\hbox{Re}\,(g|g)^-_R
\left(
2F_{R}^{*}XF_{R}-\{X,F_{R}^{*}F_{R}\}\right)+
\hbox{Re}\,(g|g)^+_R
\left(
2F_{R}XF_{R}^{*}-\{X,F_{R}F_{R}^{*}\}
\right)
$$
is a quantum Markovian generator. Notice that the factors
$\hbox{Re}\,(g|g)_{R}^{\pm}$ are $>0$ only for
$R\in {\cal S_+}$ and vanish for $R\in {\cal S_-}$.

We will also use the following notions
\begin{equation}\label{AK_kappa1}
\theta_{0,R}^{(1)}=\hbox{Re}\,(g|g)^-_R\kappa_{1,R},\qquad
\kappa_{1,R}(X)=2F_{R}^{*}XF_{R}-\{X,F_{R}^{*}F_{R}\};
\end{equation}
\begin{equation}\label{AK_kappa-1}
\theta_{0,R}^{(-1)}=\hbox{Re}\,(g|g)^+_R\kappa_{-1,R},\qquad
\kappa_{-1,R}=2F_{R}XF_{R}^{*}-\{X,F_{R}F_{R}^{*}\}.
\end{equation}

Let us prove that generators above, i.e. the $\theta$--maps satisfy some
special properties which will be crucial to prove the existence of the
solution of the flow equation. Omitting, for simplicity, the indices
$R$, we will consider all the above defined operators $\theta_{0}$,
$\theta_{1}$,  $\theta_{-1}\left(=\theta_{1}^{*}\right)$,
$\theta_{0}^{(0,\pm 1)}$, $\theta_{0}^{(\pm 1)}$,
$\kappa_{1}$, $\kappa_{-1}$,
acting on the Uniformly Hyperfinite (UHF, cf. \cite{BR}) $C^*$--algebra
${\cal B}_S$ generated by the identity and the local operators
(those acting only on a finite number of spins). On this algebra
the limits, as $\Lambda\to {Z}^d$, of all the structure maps are well
defined and their domains contain, as a common core, the unital dense
$*$-subalgebra  ${\cal B}_0$ of the local operators which is also invariant
under the square root. All our operators
(as well as the operators $F$ and $F^{*}$) map ${\cal B}_0$ into itself.
Moreover $\theta_{1}$,  $\theta_{-1}$ are nonsymmetric derivations and
$\theta_{0}^{(0,\pm 1)}$ are symmetric derivations.
From the above described properties and from (3.2.22) of \cite{BR} it follows
that $\kappa_{1}$, $\kappa_{-1}$ are (symmetric) dissipations.
But in fact $\kappa_{1}$, $\kappa_{-1}$ have a much stronger properties,
than dissipativity. We will use this property in this section without
further comments. The crucial point is the following lemma.
\medskip

\noindent
{\bf Lemma 1.}\qquad {\sl
For arbitrary local $X$, $Y$, $i=\pm1$ we have }
\begin{equation}\label{AK_stoc_der}
\kappa_{i}(XY)=
\kappa_{i}(X)Y+X\kappa_{i}(Y)+
2\theta_{-i}(X) \theta_{i}(Y)
\end{equation}

\noindent{\it Proof\/}.\qquad
Let us prove (\ref{AK_stoc_der})  for $i=-1$
(for $i=1$ the proof is analogous). Formula (\ref{AK_kappa-1}) gives
$$
\kappa_{-1}(T)=2FTF^{*} -T FF^{*} -FF^{*} T
$$
so that
$$
\kappa_{-1}(T)= [F,T]F^{*}+F[T,F^{*}]
$$
For $T=XY$ we get
$$
\kappa_{-1}(XY)=[F,XY]F^{*}+F[XY,F^{*}]=
$$
$$
=[F,X]Y F^{*} +X[F,Y]F^{*}+
F[X,F^{*}]Y+FX[Y,F^{*}]=
$$
$$
=
[F,X]F^{*}Y+X[F,Y]F +
F[X,F^{*}]Y+XF[Y,F^{*}]+2[F,X][Y,F^{*}]=
$$
$$
=\kappa_{-1}(X)Y+X\kappa_{-1}(Y)+
+2(-i)[X,F](-i)[Y,F^{*}]
$$
From (\ref{AK_minus1}), (\ref{AK_plus1}) it follows
that this is exactly (\ref{AK_stoc_der}) and that proves the lemma.

\medskip

\noindent
{\bf Theorem 1.}\qquad {\sl
For any pair of local operators $X$, $Y$, the structure maps in the
Langevin equation (\ref{AK_Langevin}) satisfy the equation
\begin{equation}\label{AK_stoc_der_com}
\theta_{\alpha}(XY)=\theta_{\alpha}(X) Y +X\theta_{\alpha}(Y)
+\sum_{\beta,\gamma}c_{\alpha}^{\beta\gamma}
\theta_{\beta}(X)\theta_{\gamma}(Y)
\end{equation}
where the structure constants $c_{\alpha}^{\beta\gamma}$
is given by the Ito table
\begin{equation}\label{AK_ito_table}
dM^{\beta}(t)dM^{\gamma}(t)=
\sum_{\alpha}c_{\alpha}^{\beta\gamma}dM^{\alpha}(t)
\end{equation}
The conjugation rules of $dM^{\alpha}(t)$ and $\theta_{\alpha}$
are connected in such a way that formula (\ref{AK_Langevin})
defines a $*$--flow ($*\circ j_t=j_t\circ *$).
}

\noindent{\it Proof\/}.\qquad
Follows from Lemma 1 and formulae (\ref{AK_minus1}), (\ref{AK_plus1}),
(\ref{AK_zero}) by direct calculation.

\medskip

The existence of the infinite volume dynamics and and its approximation by
finite volume ones, is discussed in \cite{AcKo99b}, \cite{AcKo99c}.

\section{The simplest case:
one dimensional nearest neighbor interaction}

In this section in order to compare our results with the known
results on Glauber dynamics we consider the simplest case of
one dimensional translationally invariant Hamiltonian
with the nearest neighbor interaction
$$
J_{rs}=J_{r+1,s+1},\qquad J_{rs}=J_{r,r+1}=J>0
$$
In this case for every $r$ we have 4 configurations
$\sigma(r)\in\Sigma(r)$ of nearest neighbors of the spin at $r$.
We will denote these configurations $++$, $+-$, $-+$ and $--$
(the first symbol is the orientation of the spin on the left
of $r$ and the second --- on the right).
Only the configuration $++$
will give a contribution in the stochastic limit of
interaction Hamiltonian (will lie in $\Sigma_+(r)$)
and the energies $E=E(r,\sigma)=2J$ for different $r$ will be equal.
Then (\ref{AK_fullint}) takes the form
\begin{equation}\label{AK_fullintL}
\sum_{\varepsilon,\mu=++, +-, -+, --}
F^{(\varepsilon,\mu)*}_{\Lambda}
\int dk \,\overline{g}(k)
e^{-it\left(\omega(k)-2J)\right)}a(k) + h.c.
\end{equation}
$$
F^{(\varepsilon,\mu)}_{\Lambda}=
\sum_{r\in\Lambda}F^{(\varepsilon,\mu)}_{r}
$$
$$
F^{(++)}_r=|1_{r-1}\rangle\langle 1_{r-1}|
|1_{r}\rangle\langle -1_{r}||1_{r+1}\rangle\langle 1_{r+1}|+
|-1_{r-1}\rangle\langle -1_{r-1}|
|-1_{r}\rangle\langle 1_{r}||-1_{r+1}\rangle\langle -1_{r+1}|
$$
and the other operators $F^{(\varepsilon,\mu)}_r$ are defined
correspondingly. In this case there is only one quantum white noise,
denoted $b(t,k)$, which is the stochastic limit of
$$
{1\over\lambda}\,e^{-{it\over\lambda^2}(\omega(k)-2J)}a(k)
$$
and satisfies the commutation relations
$$
[b(t,k),b^*(t',k')]=2\pi\delta(t-t')\delta(\omega(k)-2J)\delta(k-k')
$$

The Langevin equation (\ref{AK_Langevin}) takes the form
$$
dj_t(X)=\sum_{\alpha=-1,0,1}j_t\circ \theta_{\alpha}(X) dM^{\alpha}(t)
$$
where
$$
dM^{-1}(t)=dB(t),\quad \theta_{-1}(X)=-i[X,F^{(++)*}_{\Lambda}]
$$
$$
dM^{1}(t)=dB^{*}(t),\quad \theta_{1}(X) =-i [X,F^{(++)}_{\Lambda}]
$$
$$
\theta_0(X)=
\left(
\theta_{0}^{(0,-1)}+
\theta_{0}^{(0,1)}+
\theta_{0}^{(-1)}+
\theta_{0}^{(1)}
\right)(X)=
$$
$$
=\biggl(
\sum_{\varepsilon,\mu}\left(
-i \hbox{Im} \,(g|g)^-_{(\varepsilon\mu)}
[X, F_{\Lambda}^{\varepsilon\mu*} F_{\Lambda}^{\varepsilon\mu} ]
+i \hbox{Im} \,(g|g)^+_{(\varepsilon\mu)}
[X, F_{\Lambda}^{\varepsilon\mu} F_{\Lambda}^{\varepsilon\mu *}]
\right)+
$$
$$
+\hbox{Re}\,(g|g)^-_{(++)}
\left(
2F^{(++)*}_{\Lambda}X F^{(++)}_{\Lambda}-
\{X,F^{(++)*}_{\Lambda}F^{(++)}_{\Lambda}\}\right)+
$$
$$
+\hbox{Re}\,(g|g)^+_{(++)}
\left(
2F^{(++)}_{\Lambda}XF^{(++)*}_{\Lambda}-
\{X,F^{(++)}_{\Lambda}F^{(++)*}_{\Lambda}\}
\right)
$$

\bigskip

\noindent
\centerline{\large\bf ACKNOWLEDGEMENTS}

S.Kozyrev is grateful to Luigi Accardi and Centro Vito Volterra
where this work was done for kind hospitality.
This work was partially supported by INTAS 96-0698 grant.


\begin{thebibliography}{99}


\bibitem{AcFriLu}  Accardi, L., Frigerio, A., Lu, Y.G.,
''The Weak Coupling Limit  For Fermions'',
{\it Journ. Math. Phys.}\/ {\bf 32}, pp.1567--1581 (1991).

\bibitem{AcLuVo}
Accardi, L., Lu, Y.G., Volovich, I.V.,
''Interacting Fock spaces and Hilbert module extensions of the Heisenberg
commutation relations'',
Publications of IIAS, Kyoto, (1997).

\bibitem{AcLuVo99}
Accardi, L., Lu, Y.G., Volovich, I.V.,
{\it Quantum Theory and its Stochastic Limit},
Springer Verlag, 2000, to appear.

\bibitem{AcLuVo99b}
Accardi, L., Lu, Y.G., Volovich, I.V.,
''A white noise approach to classical and quantum stochastic calculus'',
Preprint of Centro Vito Volterra  N.375, Rome, July 1999.

\bibitem{AcKoVol}    Accardi, L., Kozyrev, S.V., Volovich, I.V.,
''Dynamics of dissipative two-level system in the stochastic approximation'',
{\it Phys.Rev.A}\/ {\bf 57}, N3 (1997), quant-ph/9706021.

\bibitem{AcKo99a}    Accardi, L., Kozyrev, S.V.,
''Glauber dynamics from stochastic limit'', in:
White noise analysis and related topics, Volume in honor of T.Hida,
CESNAM, Kyoto, 1999.


\bibitem{AcKo99b}    Accardi, L., Kozyrev, S.V.,
''The structure of stochastic flows'', submitted to:
{\it Infinite dimensional Analysis, Quantum Probability and related topics}\/
(1999).


\bibitem{AcKo99c}    Accardi, L., Kozyrev, S.V.,
''Stochastic dynamics of of lattice systems in stochastic limit'',
to appear in: {\it Chaos, Solitons and Fractals}\/.


\bibitem{BR}  Bratteli, O., Robinson, D.W.,
{\it Operator algebras and quantum statistical mechanics 1},
New York Heidelberg Berlin, Springer--Verlag, 1979.

\bibitem{Glauber}  Glauber, R.J.,
''Time dependent statistics of the Ising model'',
{\it J. Math. Phys.}\/ {\bf 4}, pp.294--307 (1963).

\bibitem{Liggett}  Liggett, T.M.,
{\it Interacting particle systems}\/,
Berlin, Heidelberg, New--York, Springer Verlag, 1985.

\bibitem{MajZeg96a}  Majewski, A.W., Zegarlisnki, B.,
''Quantum stochastic dynamics II'',
{\it Reviews in Mathematical Physics}\/  {\bf 8}, N5 pp.689--713 (1996).

\bibitem{Martin}  Martin, P.A., Buffet, E.,
''Dynamics of the Open BCS Model'', Preprint.

\bibitem{Matsui}  Matsui, T.,
''Markov semigroups which describe the time evolution of some
higher spin quantum models'', {\it J. Funct. Anal.}\/ {\bf 116},
pp.179--198 (1993).

\bibitem{Spitzer}  Spitzer, F.,
''Random fields and interacting particle systems'',
Proc. M.A.A. Summer Seminar, Math. Ass. Amer., Washington DC 1971.

\bibitem{Sullivan}  Sullivan, W.G.,
''Mean square relaxation times for evolution of random fields'',
{\it Commun. Math. Phys.}\/ {\bf 40}, pp.249--258 (1975).



\end{thebibliography}
\end{document}